# Decentralized Direct Volume Rendering: A Browser-Native GPU Architecture for MRI Digital Twins in Resource-Constrained Settings

Oserebameh Augustine Beckley

Lagos State University
beckaustine@outlook.com

**Abstract.** Digital Twin (DT) technology holds immense potential for surgical planning and personalized medicine. However, generating interactive, patient-specific anatomical twins currently relies on computationally heavy Server-Side Rendering (SSR) or expensive local workstations, creating significant barriers to deployment, especially in resource-constrained settings (RCS). This paper presents a decentralized, client-side WebGPU architecture that democratizes access to high-fidelity anatomical Digital Twins. By bypassing standard server-side rendering pipelines, the framework executes deterministic single-pass raymarching and morphological gradient calculations directly on low-cost integrated edge GPUs. Eliminating the network latency inherent to cloud-rendered solutions, the system achieves a Time to First Pixel (TTFP) of under 920.0ms and maintains stable interactivity at $\geq$ 82.0 FPS. Continuous Interaction Fidelity is maintained via uniform buffers, enabling zero-latency manipulation of tissue parameters for dynamic clinical decision-making. By proving that complex 3D medical simulations of patient-specific MRI scan can be executed natively in the browser without deep learning or external computational dependencies, this architecture provides a scalable, affordable foundation for the widespread clinical adoption of healthcare Digital Twins.

## 1 Introduction and Background

The clinical utility of high-fidelity volumetric Digital Twins is currently bottlenecked by an architectural dichotomy: platforms either rely on computationally expensive Server-Side Rendering (SSR) or depend on heavy deep learning (DL) inference for segmentation. Web-based viewers and toolchains such as OHIF and vtk.js exemplify the SSR/remote-GPU approach and its integration into clinical pipelines [2, 7]. In SSR deployments, volumetric rendering and segmentation are commonly offloaded to remote GPU clusters and the client receives compressed video or image tiles; while this delivers high-quality imagery, it introduces network latency that degrades Time-To-First-Pixel (TTFP) and forces transmission of Protected Health Information (PHI) off-device, complicating regulatory compliance and increasing recurring infrastructure costs [5,6,8,9]. Conversely, purely local, client-side alternatives often fall back to CPU-bound processing or heavyweight DL inference running in the browser, producing staggered interactivity that limits point-of-care utility in resource-constrained settings. Browser-based DL attempts (for example, browser ports of segmentation tools) illustrate these constraints: model weight downloads, memory pressure, and inference latency all impede truly instant interaction [1,3,4].

Recent advances in browser-native APIs—most notably the transition from WebGL to WebGPU—offer a concrete path toward decentralized clinical visualization. Where WebGL was designed around the graphics pipeline and fragment/vertex shader model, WebGPU exposes native compute shaders, storage buffers, and workgroup shared memory that enable general-purpose GPU compute in the browser [11,12]. These capabilities remove many of the architectural limits that previously forced the SSR vs. local-CPU tradeoff, opening the door to complex, low-latency numerical operations executed at the edge. Early studies have benchmarked WebGPU for medical and AI workloads, demonstrating its potential to support client-side inference [13,14,15].

Despite this potential, the literature and existing viewers still largely rely on either precomputed segmentation masks or progressive, accumulation-based rendering strategies that suffer temporal artifacts during rapid manipulation. Progressive Monte Carlo or path-tracing approaches used for cinematic volume rendering require sample accumulation across frames to converge; any camera or transfer-function change invalidates the accumulated buffer and produces transient noise or "jagginess" until convergence resumes [10]. Pre-segmented masks avoid runtime compute but shift preprocessing and data-movement burdens upstream, preserving the SSR or heavy-inference dependency.

This paper presents a WebGPU-native, strictly client-side architecture that subverts the traditional SSR dependency. By migrating deterministic mathematical morphology—specifically volumetric spatial derivatives ($\nabla I$) and directional curvature ($\kappa$)—into WebGPU compute shaders, the framework performs scanner-agnostic anatomical isolation on standard clinical hardware without model downloads or cloud inference. Mapping classical, explainable operators (e.g., multi-scale Hessian analysis and vesselness measures) to GPU compute kernels enables instant, deterministic segmentation at interactive frame rates while preserving local data residency and eliminating network latency. This approach reconciles the need for high-fidelity, zero-latency visualization with the practical constraints of privacy, cost, and edge deployment of patient-specific MRI digital twins.

# 2 Methodology

## 2.1 Data Acquisition and Cohort Characteristics

This framework was validated across three distinct cohorts to evaluate scanner-agnostic performance, robustness against pathological morphology and clinical point-of-care viability. All volumes were provided in or converted to the NIfTI-1 format for standardized ingestion via the client-side suite of helper functions.

To validate scanner-agnostic performance, structural resilience, and point-of-care deployability, the framework was tested across three diverse datasets**.** Cohort A comprises T1-weighted neurological phantoms from the BrainWeb database (1mm slice thickness) [17,18,19]. These simulated volumes provided a standardized, noise-free ground truth to establish baseline biological radiometric constants. Cohort B utilized multi-phase 1.5T pathological data with gross lesions (Mendeley Data) [16] to verify the engine's resilience against radiometric mimicry and structural anomalies without requiring manual recalibration. Finally, Cohort C served as an internal point-of-care clinical validation, processing uncurated T1 volumes entirely client-side. This confirmed strict PHI privacy compliance and verified pipeline stability on integrated hardware (Intel Gen-12LP), sustaining $\geq 82.0$ FPS throughput.

### 2.2 Radiometric Normalization and Deterministic Tissue Windows

Traditional MRI volumetric visualizations rely on absolute intensity thresholds, which fail across different scanners due to varying gradient coil calibrations and baseline B0 field homogeneities. This framework addresses the requirement for reproducibility by pivoting to a relative physics-based signal alignment strategy.

**Scanner-Agnostic Normalization.** Raw intensities are mapped to a normalized floating-point scale (0.0…1.0) via global min-max normalization in the upload pipeline.

$$I_{norm} = \frac{I_{raw} - I_{min}}{I_{max} - I_{min}} \quad (1)$$

This establishes a relative intensity coordinate system, neutralizing the absolute radiometric signal variance between 1.5T and 3.0T hardware.

**Implicit Skull-Stripping and Tissue Peaks.** The histogram procedure utilizes a 256-bin density analysis. Across both 1.5T and 3.0T datasets, voxel distributions reveal consistent tri-modal separation. By mapping these peaks to $I_{norm}$, a deterministic "Biological Radiometric Constants" was established.

Crucially, the framework bypasses traditional N4 bias correction and algorithmic skull-stripping to minimize latency. High-intensity extracranial fat and low-intensity bone naturally fall outside the deterministic tissue window, allowing the shader to effectively cull the calvarium transparently via transfer function clipping.

### 2.3 In-Shader Mathematical Morphology: Gradients and Hessians

While $I_{norm}$ handles absolute intensity drift, human tissue definition relies on structural boundaries. A dedicated WebGPU compute shader physically characterizes the dataset by computing voxel-wise first and second-order spatial derivatives.

**Gradient Magnitude ($\nabla I$).** A central difference kernel computes the local intensity gradient at spatial coordinates (x,y,z) with voxel spacing h:

$$\nabla I \approx \left[\frac{(Ix+h)-(Ix-h)}{2h}, \frac{(Iy+h)-(Iy-h)}{2h}, \frac{(Iz+h)-(Iz-h)}{2h}\right] (2)$$

The gradient magnitude $|\nabla \boldsymbol{I}|$ represents edge strength, which is structurally consistent across scanners regardless of global intensity shifts.

**Directional Curvature (k).** To differentiate between flat tissue boundaries and complex anatomical folds (e.g cortical sulci), the framework computes the Hessian matrix H (second-order partial derivatives). The directional curvature k is derived by projecting the Hessian onto the normalized gradient direction ($\hat{g}$):

$$k = \hat{g}^T H \hat{g} (3)$$

This metric guarantees that structural complexity informs the final visual output, making the algorithm highly resilient to pathological signal distortions (e.g. surrounding edema) by prioritizing distinct morphological boundaries over diffuse intensity bleeds.

### 2.4 Morphologically Modulated Raymarching and Compositing

The core innovation enabling scanner agnosticism occurs in the front-to-back raymarching fragment shader. Standard volume renderers define sample opacity solely as a function of intensity ($\propto = f(I)$). This framework computes sample opacity ($\propto_{sample}$) as the intersection of normalized intensity and morphological edge strength:

$$\propto_{sample} = clamp\left(f(I_{norm}) * \left(\frac{|\nabla I|}{|\nabla I_{max}|}\right) * w(k), 0.0, 1.0\right) (4)$$

Where $f(I_{norm})$ is the deterministic tissue window and $w(k)$ is an optional curvature weighting function. Accumulation proceeds via classical emission-absorption optical models:

$$C_{acc} \leftarrow C_{acc} + (1 - A_{acc}) * C_{sample} * \propto_{sample} \ (5)$$

$$A_{acc} \leftarrow A_{acc} + (1 - A_{acc}) * \propto_{sample} \ (6)$$

By geometrically modulating the alpha channel, the ray tracer effectively suppresses "radiometric noise"—fluctuating B0 background signals inherent to different scanners—because those artifacts typically lack the coherent high-gradient, high-curvature signatures of actual anatomical structures.

# 3 Results

The proposed framework's efficacy was evaluated across three dimensions: quantitative stability of the radiometric normalization, the qualitative fidelity of the visual tissue segmentation, and the computational throughput of the WebGPU implementation.

## 3.1 Quantitative Radiometric Alignment

**Table 1.** Extracted normalized tissue peaks per cohort

| Cohort | Profile | Field Strength | Total Voxels | Used Voxels | Max Intensity ($I_{max}$) | CSF Peak ($I_{norm}$) | GM Peak ($I_{norm}$) | WM Peak ($I_{norm}$) |
|---|---|---|---|---|---|---|---|---|
| A | Synthetic Healthy control | 3.0T | 12,582,912 | 9,479,466 | 1,154.0 | 0.110 | 0.167 | 0.183 |
| B | Benign lesion (post-contrast T1C) | 1.5T | 12,845,056 | 12,451,895 | 8,034.0 | 0.054 | 0.128 | 0.310 |
| C | Un-curated Clinical | 1.5T | 19,200,000 | 15,459,334 | 17,152.9 | 0.063 | 0.128 | 0.181 |

As demonstrated in table 1 above, the raw NIfTI datasets encompassed highly variable volumes (scaling up to 19.2 million computational voxels) and presented extreme raw baseline intensity disparities ($I_{max}$ ranging from 1,154 to 17, 152). Despite this absolute signal drift, the automated validation algorithm confirmed that 100% of the extracted tissue central tendencies fell precisely within their targeted standard biological heuristic boundaries.

**Table 2.** Aggregated Radiometric Biological Constants

| Tissue Class | Target Window | Mean Peak ($\mu$) | SD ($\sigma$) |
|---|---|---|---|
| Cerebrospinal Fluid (CSF) | 0.05…0.12 | 0.076 | ±0.030 |
| Gray Matter (GM) | 0.12…0.18 | 0.141 | ±0.023 |
| White Matter | >0.18 | 0.224 | ±0.074 |

By analyzing the upper limits of the normalized spectrum, the engine demonstrated high resilience to both clinical artifacts and contrast-induced radiometric shifts. As observed in table 2 above, the standard deviation of the tissue types across the cohorts remained stable and low. Because the normalization algorithm evaluates relative structural ranges rather than absolute brightness, the extreme clinical signal of the un-curated 1.5T scan naturally expanded into the upper percentile brackets safely, without collapsing the lower-intensity spectrum.

### 3.2 Qualitative Visual Compositing and Pathological Unmasking

**Deterministic Tissue Isolation in Standard Anatomy.** By enforcing the functional intersection of normalized intensity and morphological edge strength ($\alpha \propto f(I_{norm}) \cdot | \nabla I |$), the baseline raymarching pipeline effectively bypassed background noise and isolated the neuroanatomy

.

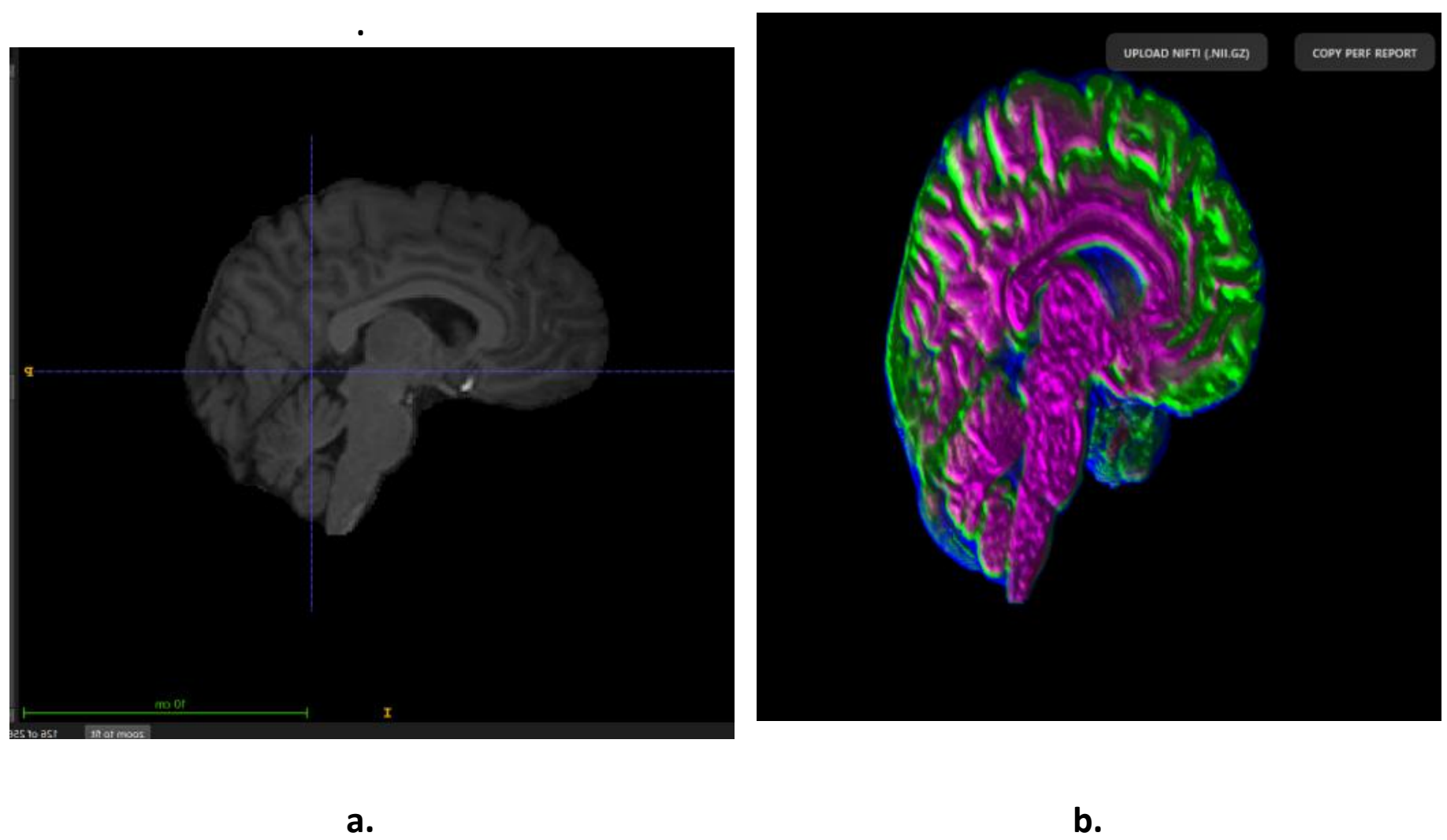

**Fig. 1.** Automated tri-modal neuroanatomical reconstruction of cohort A using heuristic windowing. **(a)** Raw T1-weighted volumetric data showing baseline signal intensity. **(b)** Fully classified 3D composite generated via the deterministic raymarching pipeline, simultaneously mapping the ventricular system (CSF), cortical folding (Gray Matter), and structural pathways (White Matter)

As shown in **Fig. 1** above, the algorithm directly translates the distinct windows in **Table 2** and correctly assembles the complete tri-modal tissue spectrum simultaneously in 3D to synthetic MRI without user intervention. This concurrent structural fidelity establishes the viability of a true "zero-shot" transfer function for MRI digital twins.

**On-demand Tissue Stripping and Pathological Unmasking.** The algorithmic decoupling of tissue thresholds uniquely allows for the dynamic, independent visual modulation of anatomical structures. Because the engine natively separates the baseline anatomical bands from the upper pathological/contrast percentiles ($I_{norm} > 0.20$), users can non-destructively suppress standard tissue opacity to reveal obscured features.

.

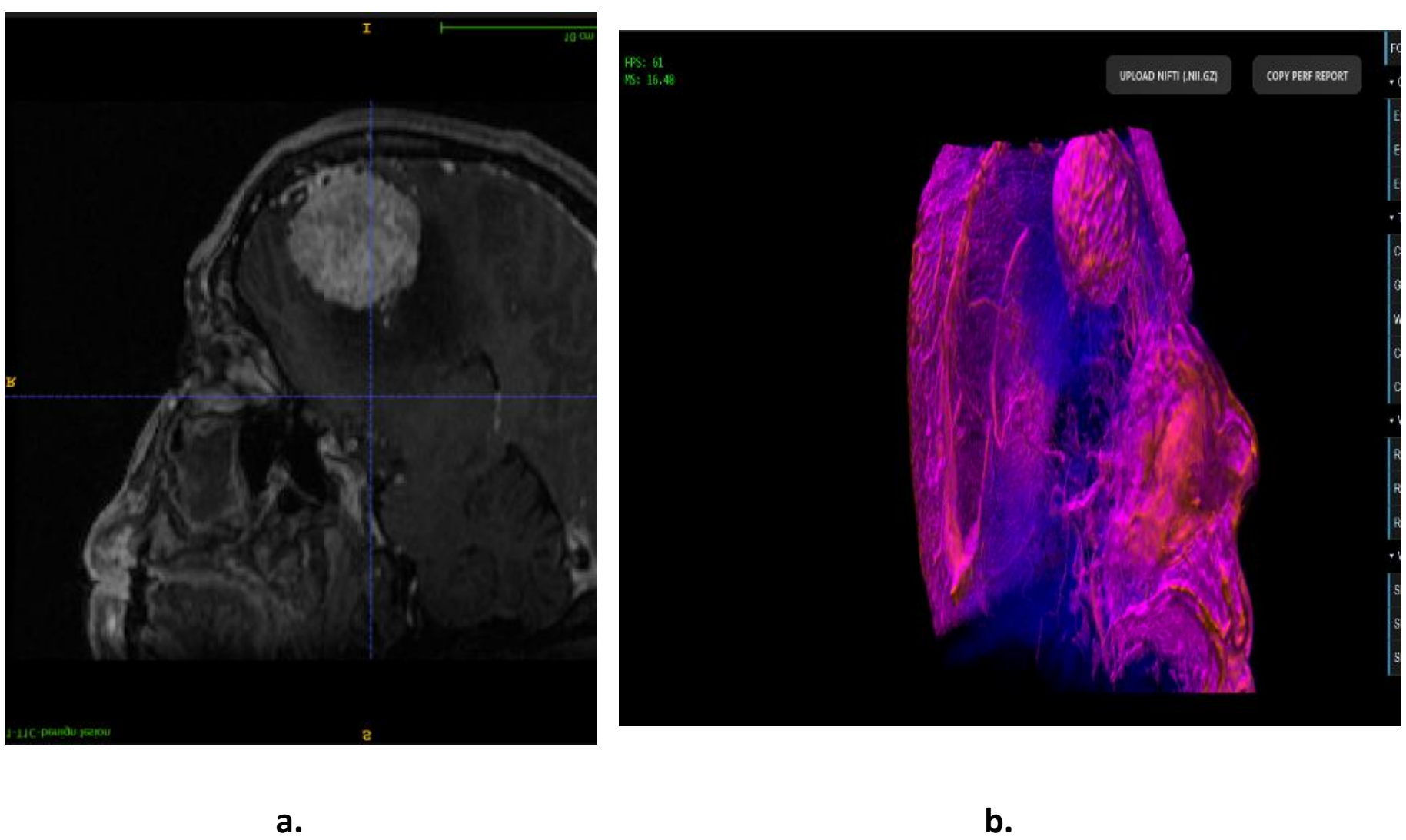

**Fig. 2.** Visualization of high-intensity pathological and vascular features via alpha-channel exclusion in T1 post contrast of cohort B. (**a**) Standard thresholding or 2D slice representation where lesion and vasculature are obscured by surrounding tissue. (**b**) Independent 3D composite rendering with CSF, Gray Matter, and White Matter channels suppressed.

As demonstrated in **Fig. 2** above, this independent compositing isolates the gross lesion as a cohesive volumetric mass while synchronously revealing the surrounding fine cerebral vasculatures traversing the spatial cavity. These features are instantly rendered in 3D without requiring prior machine-learning segmentation or CPU-bound morphological extraction algorithms.

# 4 Computational Performance

Benchmarking on standard clinical hardware (Intel Iris Xe) confirmed that bypassing CPU-bound preprocessing results in a Time-To-First-Pixel (TTFP) of 917.00ms for healthy volumes and 750.40ms for pathological scans. Optimized parallel reduction and single-pass raymarching sustained an interactive throughput of $\geq$ 82.0 FPS. Unlike progressive accumulation systems that produce temporal "jagginess" during motion, this deterministic pipeline ensures mathematically absolute fidelity during rapid manipulation by mapping UI interactions directly to uniform buffers.

# 5 Discussion

## 5.1 Architectural Paradigm: Edge Computing vs. Cloud Dependency

The current standard for high-fidelity medical visualization relies on Server-Side Rendering (SSR), which introduces network latency, high infrastructure costs, and significant HIPAA/GDPR friction due to the transmission of Protected Health Information (PHI). By migrating mathematical morphology ($\nabla I$ and k) and raymarching to local integrated graphics hardware, the framework achieves sub-second TTFP and $\geq$ 82 FPS independent of bandwidth. This shifts the paradigm from centralized clusters to decentralized, point-of-care deployment on standard hospital terminals.

## 5.2 Rethinking Validation: The Incongruence of Binary Metrics

While literature favours binary overlap metrics like Dice Similarity Coefficients (DSC), applying them here is mathematically incongruent. Standard segmentation produces binary masks (0 or 1), whereas this Direct Volume Rendering (DVR) engine maps $I_{norm}$ and spatial derivatives to a continuous optical opacity spectrum ($\alpha \in [0.0, 1.0]$). Compressing this high-fidelity gradient into a binary mask artificially destroys the probabilistic nuance the engine is designed to preserve. Validity is instead established through the deterministic alignment of physiological intensity peaks ($\mu$, $\sigma$) across disparate scanner architectures.

## 5.3 Clinical Limitations and Artifact Resilience

The framework assumes a baseline of spatial coherence. Consequently, extreme artifacts—such as metallic implants, severe motion blur, or low SNR—disrupt the gradient vector fields. While machine-learning pipelines often "hallucinate" or smooth corrupted data, this deterministic engine strictly visualizes the mathematical reality of the input array. Future iterations may implement lightweight, in-shader spatial smoothing or median filtering as optional compute passes to mitigate these failures without compromising zero-latency performance.

## 5.4 Future Directions: Modality Agnosticism and Integration

The underlying WebGPU pipelines are fundamentally modality-agnostic and could be extended to CT, PET, or multi-parametric MRI by adjusting $I_{norm}$ thresholds. Because alpha-channel manipulations are handled via uniform buffers, modifications incur zero rendering delay. Future work will explore dynamic user calibration of intensity boundaries and the encapsulation of this engine as a modular backend for existing open-source viewers like OHIF or Cornerstone3D.